\documentclass[epj,referee,nopacs]{svjour}
\usepackage{graphics}
\newcommand{\ba}{\begin{eqnarray}}
\newcommand{\ea}{\end{eqnarray}}
\newcommand{\be}{\begin{equation}}
\newcommand{\ee}{\end{equation}}
\newcommand{\bnona}{\begin{eqnarray*}}
\newcommand{\enona}{\end{eqnarray*}}

\newcommand{\MeV}{\mathrm{\ MeV}}
\newcommand{\GeV}{\mathrm{\ GeV}}

\def\ie{{\it\kern-2pt i.\kern-.5pt e.\kern3pt}}
\def\eg{{\it\kern-2pt e.\kern-.5pt g.\kern3pt}}
\title{Measurement of the DA$\Phi$NE luminosity with the KLOE detector
using large angle Bhabha scattering }
\author{The KLOE Collaboration\\
F.~Ambrosino\inst{6}, A.~Antonelli\inst{2},
M.~Antonelli\inst{2}, C.~Bacci\inst{11},
P.~Beltrame\inst{3}, G.~Bencivenni\inst{2},
S.~Bertolucci\inst{2}, C.~Bini\inst{9},
C.~Bloise\inst{2}, V.~Bocci\inst{9},
F.~Bossi\inst{2}, D.~Bowring\inst{2,13},
P.~Branchini\inst{11}, R.~Caloi\inst{9},
P.~Campana\inst{2}, G.~Capon\inst{2},
T.~Capussela\inst{6}, F.~Ceradini\inst{11},
S.~Chi\inst{2}, G.~Chiefari\inst{6},
P.~Ciambrone\inst{2}, S.~Conetti\inst{13},
E.~De~Lucia\inst{2}, A.~De~Santis\inst{9},
P.~De~Simone\inst{2}, G.~De~Zorzi\inst{9},
S.~Dell'Agnello\inst{2},
A.~Denig\inst{3,}\thanks{Corresponding author: achim.denig@iekp.fzk.de},
A.~Di~Domenico\inst{9}, C.~Di~Donato\inst{6},
S.~Di~Falco\inst{7}, B.~Di~Micco\inst{11},
A.~Doria\inst{6}, M.~Dreucci\inst{2},
G.~Felici\inst{2}, A.~Ferrari\inst{2},
M.~L.~Ferrer\inst{2}, G.~Finocchiaro\inst{2},
S.~Fiore\inst{9}, C.~Forti\inst{2},
P.~Franzini\inst{9}, C.~Gatti\inst{2},
P.~Gauzzi\inst{9}, S.~Giovannella\inst{2},
E.~Gorini\inst{4}, E.~Graziani\inst{11},
M.~Incagli\inst{7}, W.~Kluge\inst{3},
V.~Kulikov\inst{5}, F.~Lacava\inst{9},
G.~Lanfranchi\inst{2}, J.~Lee-Franzini\inst{2,12},
D.~Leone\inst{3}, M.~Martini\inst{2},
P.~Massarotti\inst{6}, W.~Mei\inst{2},
S.~Meola\inst{6}, S.~Miscetti\inst{2},
M.~Moulson\inst{2}, S.~M\"uller\inst{2},
F.~Murtas\inst{2}, M.~Napolitano\inst{6},
F.~Nguyen\inst{11,}\thanks{Corresponding author: nguyen@fis.uniroma3.it},
M.~Palutan\inst{2},
E.~Pasqualucci\inst{9}, A.~Passeri\inst{11},
V.~Patera\inst{2,8}, F.~Perfetto\inst{6},
L.~Pontecorvo\inst{9}, M.~Primavera\inst{4},
P.~Santangelo\inst{2}, E.~Santovetti\inst{10},
G.~Saracino\inst{6}, B.~Sciascia\inst{2},
A.~Sciubba\inst{2,8}, F.~Scuri\inst{7},
I.~Sfiligoi\inst{2}, T.~Spadaro\inst{2},
M.~Testa\inst{9}, L.~Tortora\inst{11},
P.~Valente\inst{9}, B.~Valeriani\inst{3},
G.~Venanzoni\inst{2}, S.~Veneziano\inst{9},
A.~Ventura\inst{4}, R.Versaci\inst{2},
G.~Xu\inst{2,1}
}
\institute{Permanent address: Institute of High Energy 
Physics of Academica Sinica, Beijing, China.\and
Laboratori Nazionali di Frascati dell'INFN, 
Frascati, Italy.\and
Institut f\"ur Experimentelle Kernphysik, 
Universit\"at Karlsruhe, Germany.\and
Dipartimento di Fisica dell'Universit\`a e Sezione INFN,
Lecce, Italy.\and
Permanent address: Institute for Theoretical 
and Experimental Physics, Moscow, Russia.\and
Dipartimento di Scienze Fisiche dell'Universit\`a 
``Federico II'' e Sezione INFN, Napoli, Italy\and
Dipartimento di Fisica dell'Universit\`a e Sezione
INFN, Pisa, Italy.\and
Dipartimento di Energetica dell'Universit\`a 
``La Sapienza'', Roma, Italy.\and
Dipartimento di Fisica dell'Universit\`a ``La Sapienza''
e Sezione INFN, Roma, Italy.\and
Dipartimento di Fisica dell'Universit\`a ``Tor Vergata''
e Sezione INFN, Roma, Italy.\and
Dipartimento di Fisica dell'Universit\`a ``Roma Tre''
e Sezione INFN, Roma, Italy.\and
Physics Department, State University of New 
York at Stony Brook, USA.\and
Physics Department, University of Virginia, USA.
}
\date{Received: date / Revised version: date}
\abstract{
We describe the method of measuring the integrated luminosity of the
$e^+e^-$ collider DA$\Phi$NE, the Frascati $\phi-$factory. The
measurement is done with the KLOE detector selecting large angle
Bhabha scattering events and normalizing them to the effective cross section. 
The $e^+e^- \to e^+e^-(\gamma)$ cross
section is calculated using different event generators which account
for the $\mathcal{O}(\alpha)$ radiative initial and final
state corrections, and the
$\phi$ resonance contribution.
The accuracy of the measurement is 0.6\%,  where 0.3\% comes from systematic
errors related to the event counting and 0.5\% from theoretical evaluations
of the cross section.
\PACS{
      {PACS-key}{discribing text of that key}   \and
      {PACS-key}{discribing text of that key}
     } 
} 
\titlerunning{Measurement of the DA$\Phi$NE Luminosity}
\authorrunning{F.~Ambrosino {\it et al.}}
\begin{document}
\hugehead
\maketitle
\section{Introduction}
\label{intro}
For an accurate measurement of the cross section
of an $e^+e^-$ annihilation 
process, the precise knowledge of the collider
luminosity is required.
The luminosity depends on three factors:
beam-beam crossing
frequency, beam currents and the beam overlap area in the
crossing region. However, the last quantity is difficult to
determine accurately from the collider optics.
Thus, experiments prefer to determine the luminosity
by the counting rate of well
selected events whose cross section is known with good precision.
Since the advent of low luminosity $e^+e^-$ colliders, a great
effort was devoted to obtaining good precision in the cross section
of electromagnetic processes, extending the pioneer work of the
earlier days~\cite{TheBible}. At the $e^+e^-$ colliders,
working in the range $1\GeV<\sqrt{s}<3\GeV$, such as ACO at Orsay,
VEPP-II at Novosibirsk, and Adone at Frascati, the
luminosity measurement was based on small angle Bhabha scattering,
or single and double $e^+e^-$ bremsstrahlung~\cite{SABS,BREMS},
thanks to the high statistics. The electromagnetic cross
sections scale as 1/s, while elastic
$e^+e^-$ scattering has a steep dependence on the polar angle,
$\sim 1/\theta^3$, thus providing high rate for small values of
$\theta$. At low and intermediate energy high-luminosity meson
factories, the small polar angle region is difficult
to access for the presence of the low-beta insertions close to the
beam crossing region, while wide angle Bhabha scattering produce a
large counting rate and can be exploited for a precise
measurement of the luminosity.

We have measured the luminosity counting the number
of large angle Bhabha scattering events and 
normalizing this number to the effective Bhabha cross section
$\sigma_{\rm eff}$:
\begin{equation}
\int\!\mathcal{L}\,{\rm d} t = \frac{N_{\rm
obs}-N_{\rm bkg}}{\sigma_{\rm eff}}~.
\label{eq:1}
\end{equation}
The effective cross section is evaluated by inserting into the detector
simulation different event generators
which include radiative corrections at a high level of precision.
In eq.(\ref{eq:1}) the number of background events, $N_{\rm bkg}$,
is determined and
subtracted from the observed events, $N_{\rm obs}$.

The main advantages of this method are:
\begin{enumerate}
\renewcommand{\labelenumi}{\roman{enumi})}
\item high theoretical accuracy by which the cross section could
be calculated;
\item clean event topology of the signal and
small amount of background;
\item large statistics: for
$\sigma_{\rm eff} \sim430$ nbarn in
$45^\circ<\theta_{e}<135^\circ$, even at the lowest luminosities
obtained in the data taking period, the statistical error
$\delta\mathcal{L}/\mathcal{L}\sim0.3\%$ is reached in
about two hours of data taking.
\end{enumerate}
In the following we describe the luminosity measurement using
large angle Bhabha scattering. The on-line measurement, with 5\%
accuracy, was used to provide a fast feedback to DA$\Phi$NE. The
off-line analysis which is described in this paper 
reaches a precision of 0.6\%, dominated by the
uncertainty quoted at present, in the calculation
of the Bhabha cross section. 
A high precision on $\mathcal{L}$ is particularly useful in the
KLOE measurement of the hadronic cross section~\cite{kloeppg}.

\begin{table}
\begin{center}
\caption{Main parameters of the DA$\Phi$NE
 beams during the operation in year 2001.}
\label{tab:1}
\begin{tabular}{| l  c |}
\noalign{\smallskip}\hline\noalign{\smallskip}
number of bunches ($e^+$, $e^-$) & 49, 49 \\
current per beam (A) & 0.7, 1.0 \\
beam crossing period (ns) & 5.43 \\
beam width at crossing: &  \\
$\sigma_{x}$ (mm) (horizontal) & 2 mm \\
$\sigma_{y}$ (mm) (vertical) & 20 $\mu$m \\
$\sigma_{z}$ (mm) & 3 cm \\
average luminosity  & 3 $\times10^{31}$ cm$^{-2}$ s$^{-1}$ \\
luminosity lifetime (min) & 30 \\
\noalign{\smallskip}\hline
\end{tabular}
\end{center}
\end{table}

\section{The DA$\Phi$NE collider}
\label{sec:1}
The DA$\Phi$NE $e^+e^-$ collider~\cite{Guiducci:DAPHNE} is
designed to run at high luminosity in the energy region corresponding
to the resonance $\phi(1020)$.
It consists of two independent electron and positron rings of 98 m length with beams
that cross at two interaction regions
with angle of $\sim$ 25 mrad.
DA$\Phi$NE was commissioned in 1999 and since 2000 was working
with increasing luminosity for three experiments: KLOE, DEAR and
FINUDA.

\begin{figure}
\begin{center}
\resizebox{0.9\columnwidth}{!}{%
\includegraphics{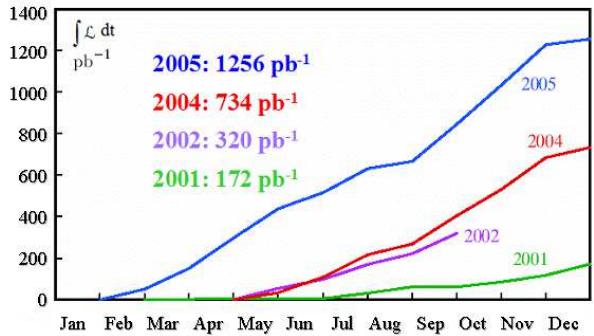}}
\caption{Luminosity collected by the KLOE experiment from 2001
to 2005.}
\label{fig:1}       
\end{center}
\end{figure}

At the end of 2005 KLOE has
collected an integrated luminosity of $\sim$ 2.5 fb$^{-1}$, as shown in
Fig.~\ref{fig:1}. The
measurement presented here only refers to the data taken during 2001,
although KLOE will use the same luminosity method for the remaining data set.
 
DA$\Phi$NE works in the "topping up" mode,
injecting beams with a frequency of about three fillings
per hour while the KLOE
experiment is continuously taking data.
The main beam parameters are presented
in Table~\ref{tab:1}.

In the KLOE interaction region (IR) electron-positron beams cross
with a small transverse momentum in the horizontal plane $p_{T}
\sim$~13 MeV/c. 
The longitudinal and horizontal width of the
beam-beam collision region is $\Delta z \sim 12$ mm and $\Delta x
\sim 1.2$ mm respectively. All these quantities are measured run-by-run
with very good accuracy ($\sigma_{p_{T}} \simeq 100$ keV,
$\sigma_{\Delta z} \simeq 0.1$ mm, $\sigma_{\Delta x} \simeq 0.05$ mm)
detecting large angle Bhabha events as it will be
explained in the following.
The beam energy spread is $(0.302\pm0.001)$ MeV, as determined
from $\phi\to K_L K_S$ decays.

\section{The KLOE detector}
\label{sec:2}
\begin{figure}
\begin{center}
\resizebox{0.75\columnwidth}{!}{%
\includegraphics{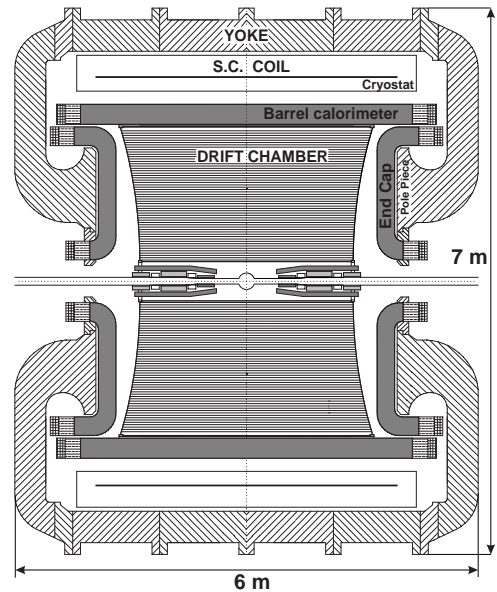}}
\caption{Cross view of the KLOE detector.}
\label{fig:2}       
\end{center}
\end{figure}

A cross view of the detector is shown in Figure~\ref{fig:2}. It
consists of a large volume drift chamber (DC), a fine grained
lead-scintillating fiber sampling calorimeter (EMC) both immersed
in a uniform magnetic field of 0.52 T parallel to the beam
bisectrix, which is taken as the axis of our coordinate system.
The beam pipe around the IR has a spherical shape of
10 cm radius. Three low-beta quadrupoles on either side, at a
distance of 50 cm from the IR, fill the space between the beam
pipe and the DC inner wall. Two small lead-scintillating tile
calorimeters are wrapped around the quadrupoles to complete the
EMC hermeticity.

The drift chamber~\cite{Adinolfi:2002DC}, 4 m in diameter and 3.3
m long, is made of 58 concentric rings of drift cells arranged in
a full-stereo geometry and is filled with a low Z
gas mixture (90\%He$-$10\%$i$-C$_{4}$H$_{10}$). Particle trajectories
are measured with a space resolution of $\sigma_{xy} \simeq$ 0.15
mm and $\sigma_{z} \simeq$ 2 mm; the transverse momentum
resolution for long tracks is $\sigma(p_{T})/p_{T} \simeq$ 0.3\%
and the primary vertex is reconstructed with a space resolution
$\simeq$ 2 mm. Electrons emitted at large angle have a track
length greater than 1.5 m and are measured with more than
50 points.

The calorimeter~\cite{Adinolfi:2002EMC} is divided in a barrel
($45^\circ < \theta < 135^\circ$) and two end-caps. It is
segmented in depth in five layers, the first four of
$\sim 3 X_0$ each, and the fifth of $\sim 4 X_0$. The barrel is divided
in 24 sectors, each sector having 5$\times$12 calorimeter cells of
4$\times$4 cm$^{2}$ read out by photomultipliers at both ends to
measure the arrival time of particles and to reconstruct the space
coordinates.
Calorimeter clusters are
reconstructed grouping together energy deposits close in space and
time. For each cluster, $E_{cl}$ is the sum of the cell energies,
the time $t_{cl}$ and position $\vec{r}_{cl}$ are calculated as
energy-weighted averages. The energy, time and position
resolutions are $\sigma_{E}/E = 0.057/\sqrt{E(\mathrm{GeV})}$,
$\sigma_{t}$ = 54 ps/$\sqrt{E(\mathrm{GeV})} \oplus 50$ ps,
$\sigma_{xy}\sim 1.3$ cm, and $\sigma_z \sim$ 1 cm$/\sqrt{E(\mathrm{GeV})}$.

The trigger~\cite{Ambrosino:2004TRIG} is based on the information
from the calorimeter and the drift chamber. The EMC trigger
requires two distinct energy deposits above threshold (E $>$ 50 MeV in the
barrel and E $>$ 150 MeV in the end-caps). The DC trigger is based
on the number of drift cells that recorded a hit and on their
topology. Recognition and rejection of cosmic ray events is also
done at trigger level: events with two energy deposits above 30
MeV in the fifth calorimeter layer are vetoed. The trigger 
is synchronized with the DA$\Phi$NE RF divided by 4, $T_{sync}$
= 10.86 ns, with an accuracy of 50 ps. The time of the bunch
crossing producing an event is determined after event
reconstruction.
\begin{figure*}
\begin{center}
\mbox{
\resizebox{0.85\columnwidth}{0.83\columnwidth}{
\includegraphics{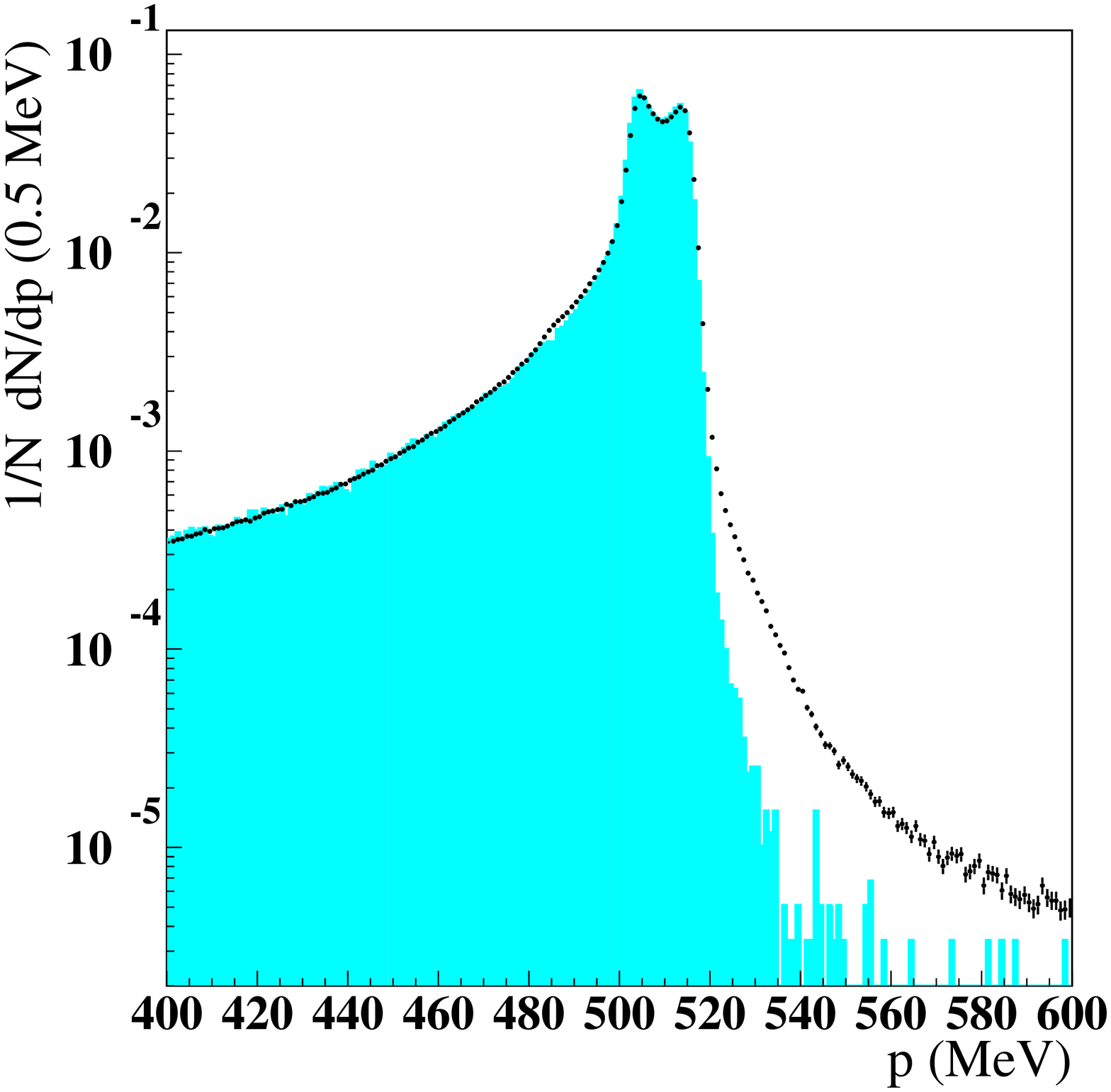}}
\hspace*{0.5cm}
\resizebox{0.85\columnwidth}{0.83\columnwidth}{
\includegraphics{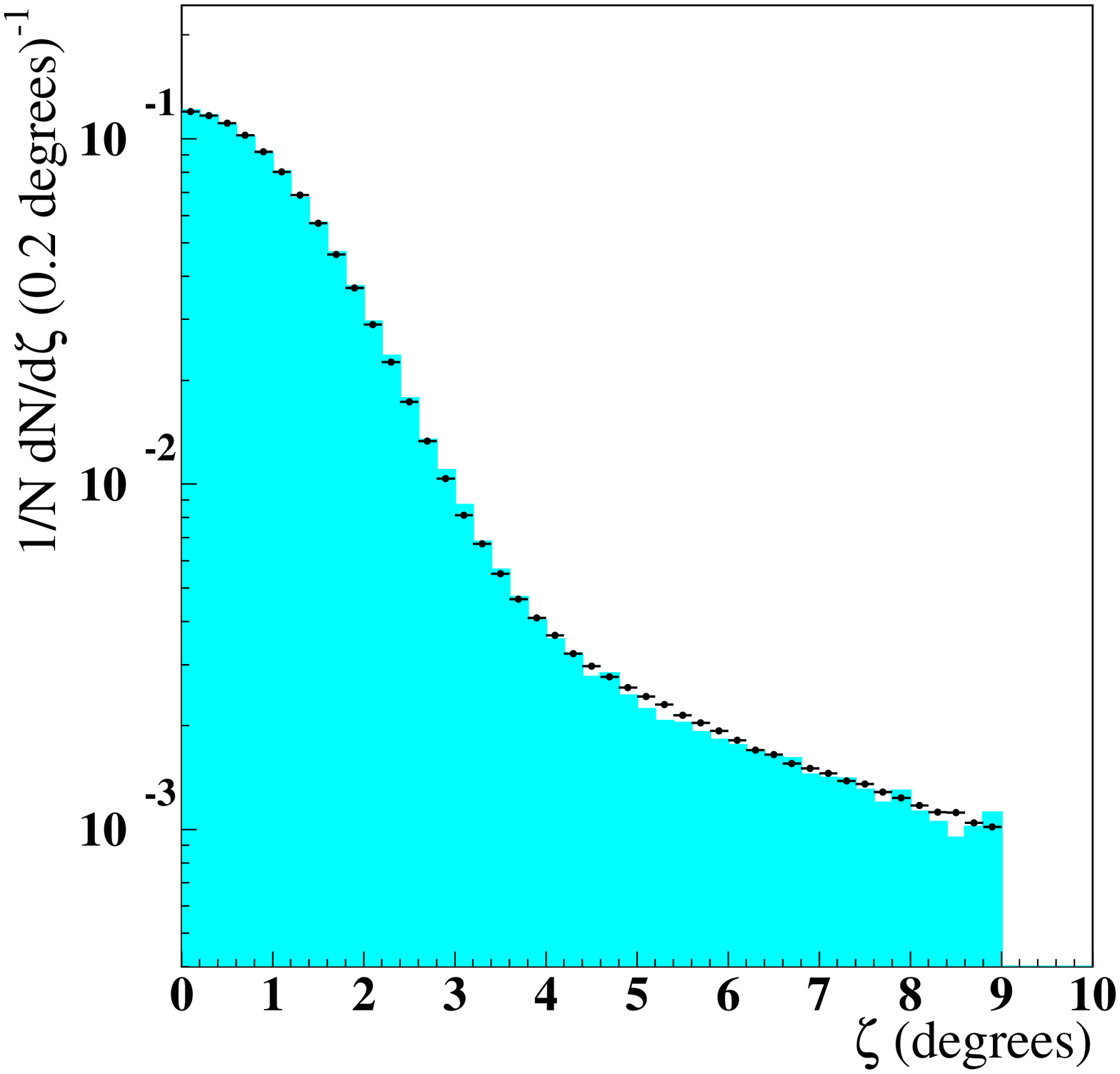}}
}
\caption{Comparison between data
(points) and Monte Carlo (histogram) distributions for the track
momentum $p$ (left) and for the energy clusters acollinearity
$\zeta$ (right).}
\label{fig:3}
\end{center}
\end{figure*}

\section{The selection of Bhabha scattering events}
\label{sec:3}
An on-line filter selects Large Angle Bhabha (\texttt{LAB})
events using only calorimeter information, to minimize
the CPU time necessary for the whole event reconstruction;
while a more refined off-line analysis is done selecting
Very Large Angle Bhabha (\texttt{VLAB}) events, tightening
the acceptance cuts and including also the tracking information.

The \texttt{LAB} selection is based on the following requirements:
\begin{enumerate}
\renewcommand{\labelenumi}{\arabic{enumi})}
\item at least two calorimeter clusters with energy
$300\MeV<E_{cl}<800\MeV$;
\item the two clusters with the minimum polar angle acollinearity
$\zeta=|\theta_{cl1}+\theta_{cl2}-180^\circ|$ are chosen
and $\zeta_{\rm min} < 10^\circ$;
\item the time cut $|t_{cl1}-t_{cl2}|<4\mbox{ ns}$, for the two clusters;
\item both clusters with $45^\circ<\theta_{cl}<135^\circ$;
\item $\cos\alpha>-0.975$, where $\cos\alpha = \vec{r}_{cl1}
\cdot \vec{r}_{cl2}/|\vec{r}_{cl1}|\ |\vec{r}_{cl2}|$, $\vec{r}$
being the cluster position (this cut is
introduced to reject $e^+\,e^-\to\gamma\,\gamma$ events, which have
a back-to-back topology);
\item the presence of at least $50$ DC hits in the event.
\end{enumerate}
The precision with which \texttt{LAB} events are selected is about
1\% and is limited by the energy resolution of the calorimeter
($\sigma_E \simeq$ 40 MeV for $E_{cl}$ = 510 MeV).
By adding information from the tracking chamber, the precision is
considerably improved and the background of $\pi^+\pi^-$ and
$\mu^+\mu^-$ events ($1.2\%$ contamination at this level) is
further reduced.

In the \texttt{VLAB} selection
the tracking information gives the momentum measurement and the charge
assignment, while the information on polar angles
is still taken from the EMC clusters. There is no need to use the
tracks for the angular information since calorimeter clusters and
tracks have similar angular resolutions ($\sigma_{\theta} \simeq
1^\circ$) and hence no further systematic uncertainty is introduced. The
selection cuts of \texttt{VLAB} events are slightly tighter than
in the \texttt{LAB} selection; the event must satisfy the
following requirements:
\begin{enumerate}
\renewcommand{\labelenumi}{\arabic{enumi})}
\item for the two tracks with the largest number of associated
hits, the point-of-closest-approach to the origin
(PCA) must be within $(x_{pca}^2+y_{pca}^2)^{1/2} < 7.5$ cm and
$|z_{pca}|<15$ cm;
\item the two tracks must have opposite curvature;
\item both tracks must have momentum $p \ge $ 400 MeV;
\item the two EMC clusters selected by the \texttt{LAB} filter
must have polar angle $55^\circ < \theta_{cl} < 125^\circ$;
\item the cut on the polar angle acollinearity for the two \texttt{LAB}
clusters is further tightened to $\zeta < 9^\circ$.
\end{enumerate}

\section{Evaluation of efficiencies}
\label{sec:5}
The effective \texttt{VLAB} cross section is obtained from Monte
Carlo, it is therefore important to check that the resolution of
the variables and the efficiency of the selection are well
reproduced by the detector simulation, and to correct for any
mismatch between data and Monte Carlo. In particular a difference
in the resolution of the kinematic variables can give rise to
systematic effects at the borders of the chosen phase space. The
studies presented here refer to the data sample collected in 2001
because these data have been used for the measurement of the pion
form factor~\cite{kloeppg}. Since 2002 the cosmic ray veto has
been improved and a smaller systematic error from this effect
should be accounted for in later data.

\subsection{Angular acceptance}
\label{subsec:3}
\begin{figure}
\begin{center}
\resizebox{0.95\columnwidth}{!}{%
\includegraphics{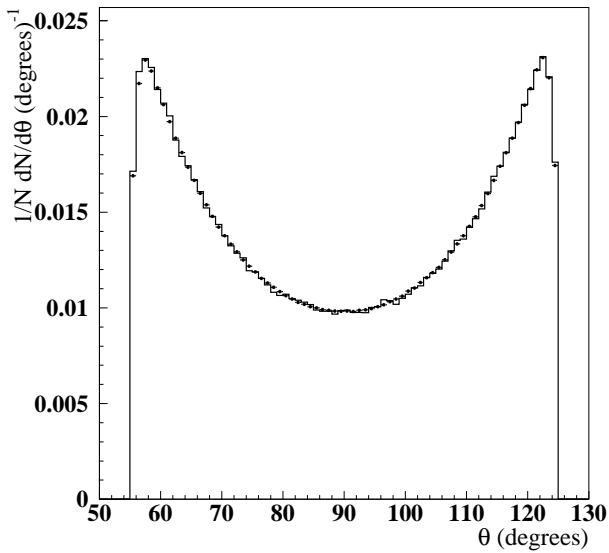}
}
\vspace*{-0.3cm}
\caption{Comparison between the data (points)
and Monte Carlo (histogram) distributions of energy clusters polar angle
normalized to the same number of events.}
\label{fig:4}
\end{center}
\end{figure}
Figure~\ref{fig:3} shows the comparison of data with the
Monte Carlo simulation~\footnote{Hereafter we
will implicitly refer to the event generator \texttt{Babayaga}, Version 3.5.
This and other generators will be discussed in Sec.~\ref{sec:4}.}
for the distributions of momentum, $p$, and acollinearity, $\zeta$.
In both variables, the cuts $p > 400\MeV$ and $\zeta < 9^\circ$ occur
where the agreement is very good.
Furthermore, both variables are cut in a region
where eventual resolution mismatches between data and Monte Carlo
have small effects, because far from the bulk of the distribution.
Thus, any systematic error from the cuts on $p$ and $\zeta$ is
considered as negligible. The difference observed at high particle
momenta is due to non-Gaussian tails in the DC reconstruction
which are not simulated in Monte Carlo. However, since no
high-momentum cut is applied in the selection, no systematic
effect arises from this small difference.

The situation is different for the requirement on the polar angle,
$\theta$. In this case the cut $55^\circ < \theta <125^\circ$ is
applied in a region with a steep rise of the distribution, with
a priori possible large systematic effects.
Figure~\ref{fig:4} shows the comparison of the data and the
simulated distribution, both normalized to the total number of
events: the overall agreement is very good. The polar angle
resolution from the measurement of the calorimeter clusters is
$\sim1^\circ$, therefore a net gain or loss of events due to a
systematic difference between the polar angle resolution in data
and Monte Carlo can only occur in the bins close to the borders.
To evaluate the effect, the relative difference between data and
Monte Carlo is computed in the border intervals ($55^\circ < \theta <
65^\circ$, $115^\circ < \theta <125^\circ$), after normalizing
the number of Monte Carlo events, $N_{MC}$, to coincide 
with the number of data events, $N_{data}$, in the central region
($65^\circ < \theta < 115^\circ$):
the value $(N_{data} - N_{MC})/N_{data} = (-0.25 \pm 0.03)\%$ is used both
as the relative correction to the effective cross section and as
systematic uncertainty on the angular acceptance. This
estimate is confirmed by computing the relative variation of the
luminosity as a function of the value of the cut in polar angle,
$\theta_{cut}$:
\bnona
 \frac{\Delta\mathcal{L}}{\mathcal{L}} & = &
 \frac{N_\mathtt{VLAB}
 (\theta_{\rm cut} < \theta < 180^\circ - \theta_{\rm cut})}
 {N_\mathtt{VLAB}(55^\circ < \theta < 125^\circ)} \\
 & - &
 \frac{\sigma_{\rm eff}
 (\theta_{\rm cut} < \theta < 180^\circ - \theta_{\rm cut})}
 {\sigma_{\rm eff}(55^\circ < \theta <125^\circ)}
\enona
The behaviour of $\Delta\mathcal{L}/\mathcal{L}$ as a function of
$\theta_{\rm cut}$ shows that, in a 5$^\circ$ range,
the relative variation is
$\Delta\mathcal{L}/\mathcal{L} = ^{+0.003}_{-0.002}$,
consistent with the quoted systematic error.

\subsection{Tracking efficiency}
\label{subsec:4}
To evaluate the tracking
efficiency we use \texttt{LAB} events because no tracking
information is required in selecting this sample.
Thus, we select events with
a tagging track having $p_{\rm tag}>400\MeV$
and associated to one
of the two \texttt{LAB} clusters.

In this subsample, we define the
tracking efficiency $\epsilon_{track}$
as the fraction of events which fulfil the
following requirements:
\begin{enumerate}
\renewcommand{\labelenumi}{\arabic{enumi})}
\item at least a second track associated to the origin (as
defined above), this track must be one of the two with the largest
number of associated hits;
\item the track must have momentum $p_{2} > 400$ MeV and
curvature opposite to the tagging track;
\item the distance $d$ between the first hits of tagging and
candidate track must be larger than 50 cm.
\end{enumerate}
We have verified that varying the values for $d$,
$p_{\rm tag}$, $p_{2}$ the tracking efficiency
$\epsilon_{track}$ is stable and we find that the efficiency for
data and Monte Carlo are:
\bnona
 \epsilon_{track}^{data} &=& (99.824 \pm 0.005)\% \\
   \epsilon_{track}^{MC} &=& (99.764 \pm 0.011)\%
\enona
where the errors are statistical. The relative difference
$\Delta\epsilon_{track}=(6.0 \pm 1.2)\times10^{-4}$ is taken as the
systematic uncertainty due to the tracking efficiency.

\subsection{Cluster efficiency}
\label{subsec:5}
To evaluate the cluster efficiency
we select a subsample based on the tracking information.
Wa ask for two and only two tracks with the following requirements:
\begin{itemize}
\renewcommand{\labelitemi}{--}
\item the two tracks are connected to one and only one vertex located at $|\vec{r}| < 5$ cm;
\item both tracks are emitted at polar angle $50^\circ < \theta <
130^\circ$, where $\theta$ is measured at the vertex position;
\item both tracks fulfil the same requirements on the radial
position of their first hit (\emph{fh}) and last hit (\emph{lh})
in the DC: $(x_{fh}^2+y_{fh}^2)^{1/2} < 40$ cm and
$(x_{lh}^2+y_{lh}^2)^{1/2} > 180$ cm (these requirements exclude
splitted tracks);
\item the electron-positron invariant mass, $M_{e e}$, must be
in the range $1017.5\MeV/c^2< M_{e e} < 1021.5\MeV/c^2$;
\item the track mass, $m_{trk}$, defined as the mass
associated to the momenta $p_{1}$ and $p_{2}$ under the hypothesis
of a final state of two charged particles of the same mass and one
photon, should be smaller than 90 MeV,
\[ \sqrt{|\vec{p}_1|^2 + m_{trk}^2} +
   \sqrt{|\vec{p}_2|^2 + m_{trk}^2} +
   |\vec{p}_1 + \vec{p}_2 - \vec{p}_b| = \sqrt{s} \]
here $\vec{p}_b$ is the average beam-beam transverse momentum
measured run by run.
\end{itemize}
The last two cuts efficiently remove the background from
$\mu^+\mu^-$ and $\pi^+\pi^-$ events. We then look for two
calorimeter clusters satisfying the requirements:
\begin{enumerate}
\renewcommand{\labelenumi}{\arabic{enumi})}
\item $|\vec{\rho}_{lh} - \vec{\rho}_{cl}| < \Delta\rho = 40$ cm,
$\vec{\rho}_{cl}$ being the position of the cluster in the $x$-$y$
plane; this defines the cluster to track association;

\item $|p - E_{cl}| < \Delta E = 210\MeV$;

\item $|t_{cl1}-t_{cl2}|<\Delta t=4$ ns.
\end{enumerate}
The cluster efficiency $\epsilon_{cluster}$ is defined as the
fraction of events in the control sample which fulfil the
requirements. We have verified that varying the values for
$\Delta\rho$, $\Delta t$ and $\Delta E$ the efficiency is stable
and we find that the values for data and Monte Carlo are in good
agreement:
\bnona
 \epsilon_{cluster}^{data} &=& (99.58 \pm 0.11) \% \\
 \epsilon_{cluster}^{MC} &=& (99.65\pm 0.02)\%
\enona
where the errors are statistical.

To evaluate the systematic uncertainty,
the inefficiency in data has been studied
and understood as due to cluster splitting, in which
a cluster is split into two clusters
neither of them surviving the lower energy cut.
A correction of $\delta_{split}=(0.135 \pm
0.007)\%$
is applied to $\epsilon_{cluster}^{MC}$, and
the following difference between data 
and Monte Carlo is obtained: 
$$\Delta\epsilon_{cluster} = \epsilon_{cluster}^{data} -\epsilon_{cluster}^{MC}
= (0.07 \pm 0.11)\%$$
Since both data and Monte Carlo agree within statistical errors,
we take the value $0.11\%$ as the systematic
uncertainty in the cluster efficiency.

\subsection{Background}
\label{subsec:6}
\begin{figure}
\begin{center}
\resizebox{0.98\columnwidth}{0.95\columnwidth}{%
\includegraphics{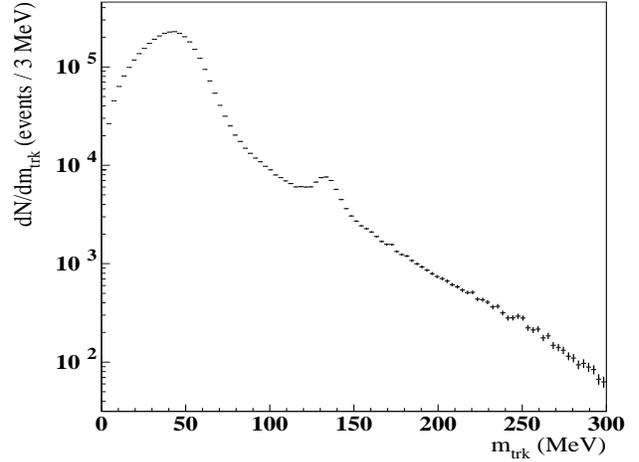}
}
\vspace*{-0.3cm}
\caption{$m_{trk}$ distribution for a
sample of \texttt{VLAB} events.}
\label{fig:5}
\end{center}
\end{figure}

Given the cut on the
track momentum, $p > 400\MeV$, the only relevant background
processes are $e^+ e^- \to \mu^+ \mu^-$ and
$e^+\,e^-\to\pi^+\,\pi^-$. The estimate of this background is
based on the track mass variable, $m_{trk}$.
Figure~\ref{fig:5} shows the $m_{trk}$ distribution for a
sample of \texttt{VLAB} events: besides Bhabha scattering events,
clustered at low values of $m_{trk}$, the only significant
structure is the peak associated with $\pi^+\,\pi^-$ events around
$m_{trk} \simeq 136\MeV$. There is no evidence for background from
$\mu^+ \mu^-$ events because of the lower cross section
and of the smaller efficiency to release clusters with $E>300\MeV$.
We have used two methods to measure the amount of
background: the first consists in fitting the track mass
distribution, while in the second method we use particle
identification based on time of flight method for
discriminating pions (muons) from electrons, called \texttt{PID}
function in the following.
\begin{enumerate}
\item The \texttt{PID} function exploits the time of
flight and the different
shape of energy deposits in the calorimeter layers,
of clusters associated to tracks to discriminate pions
(muons) from electrons on an event-by-event basis~\cite{kloeppg}.
The fraction of background is evaluated from events in which at
least one track has been identified by the
\texttt{PID} function as a pion,
\[ \frac{N_{\mathrm{bkg}}}{N_{\texttt{VLAB}}} = (0.623 \pm 0.015) \% \]
\item The $m_{trk}$ distribution from Monte Carlo
is well described by an exponential function in the
range 100 MeV $< m_{trk} <$ 170 MeV. We fit the $m_{trk}$
distribution in this range with a Gaussian (background) plus an
exponential function (signal): 
the relative amount of background is 
$$ \frac{N_{bkg}}{N_{\texttt{VLAB}}} = (0.54\pm 0.13)\% ~,$$
\end{enumerate}
Since the previous results are in agreement within statistical errors,
we take the value 0.13\% as the systematic uncertainty
and $N_{\mathrm{bkg}}/N_{\texttt{VLAB}}=0.62\%$
as the background contamination in \texttt{VLAB} event sample.

\subsection{Cosmic veto}
\label{subsec:7}
Cosmic ray events are vetoed at the trigger level, but a fraction of these
events is flagged and recorded for calibration with a
downscale factor of 5. Applying the \texttt{VLAB} selection on
the downscaled events, we estimate the total fraction of
\texttt{VLAB} events lost due the trigger veto directly
from data. The effect is
stable in time and an average correction of $(0.40 \pm 0.03)\%$
has been applied to the effective cross section evaluated
with Monte Carlo.

\section{Evaluation of systematic effects}
The effects on the acceptance and efficiency
discussed so far do not show variations in time and therefore
average corrections were applied to the whole data set. Other effects depend
on the actual run conditions and there was need to determine the
corrections on a run-by-run basis. As will be shown, these time
dependent effects are very small. In particular
they are related to energy calibration and to
variations in the center of mass energy.

\subsection{Calorimeter energy calibration}
\label{subsec:8}
We have studied the effect of a variation of the calorimeter
energy scale on the \texttt{LAB} selection, which requires two
energy clusters in the interval $300-800$ MeV and we have computed
the effect of these variations on the \texttt{VLAB} selection.
\begin{figure}
\begin{flushleft}
\resizebox{1.13\columnwidth}{!}{%
\includegraphics{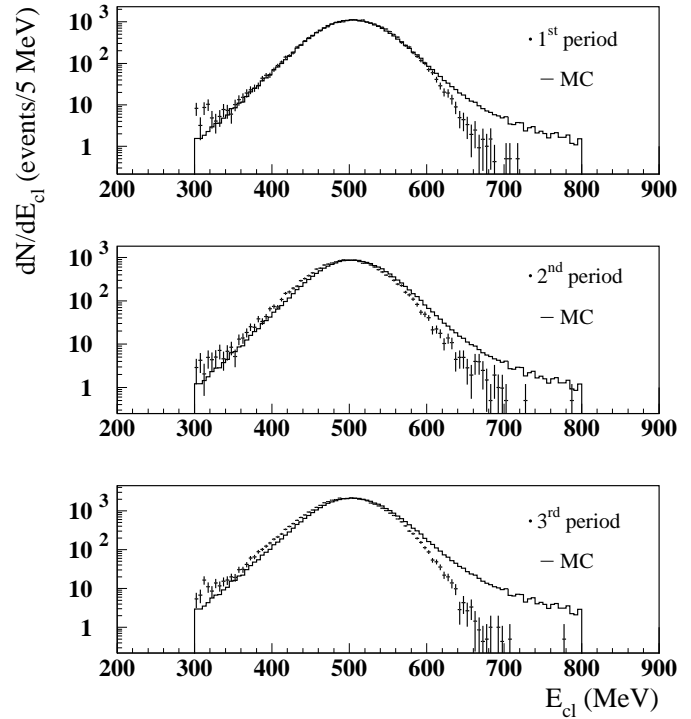}
}
\caption{Distribution of the cluster
energy of \texttt{VLAB} events for three runs (points),
one from each of the three different periods of data taking,
compared with Monte Carlo (solid line).}
\label{fig:7}
\end{flushleft}
\end{figure}
The 2001 data sample consists of three periods,
where the EMC energy scale changed
by 1\%, for an energy resolution of 8\%.
For each period, the $E_{cl}$ distribution is
evaluated and compared with the Monte Carlo distribution
(see Figure~\ref{fig:7}). We observe that i) the Monte Carlo
overestimates the high energy tail of the distribution, and ii)
there are systematic shifts in the $E_{cl}$ mean value.

We have calculated the systematic effect due to overestimating the
high energy tail by extrapolating the Monte Carlo distribution
above $800$ MeV. The relative difference between
data and Monte Carlo amounts to $\Delta E_{tail}=(6.1\pm1.6)\times10^{-4}$.
The value $\Delta E_{tail}$ is taken as the systematic uncertainty
due to different high energy tails between data and Monte Carlo.

The effect of the shift of the energy distribution between data
and Monte Carlo has been estimated by taking the difference
between the mean values of the distributions as a measurement of the
shift and by adding (subtracting) the events, which are gained
(lost) according to this shift. The run-by-run weighted average 
is $\Delta E_{calib}=(6 \pm 2)\times10^{-4}$, and it has been
considered as the relative systematic error due to variations
in the calorimeter energy calibration. Furthermore we have checked
that, aside from a coherent shift, the shape of the cluster energy
distribution in \texttt{VLAB} events is the same in the three periods
(see Figure~\ref{fig:7}).

The overall systematic error due to the calorimeter energy calibration is
\[ \Delta E_{cl} = \Delta E_{tail}\oplus\Delta E_{calib} =
   8.6\times10^{-4} \]
Since the two effects tend to compensate, no correction was
applied to the luminosity measurement.

\subsection{Center of mass energy}
\label{subsec:9}
\begin{figure}
\begin{center}
\resizebox{1.02\columnwidth}{!}{%
\includegraphics*[126,136][446,368]{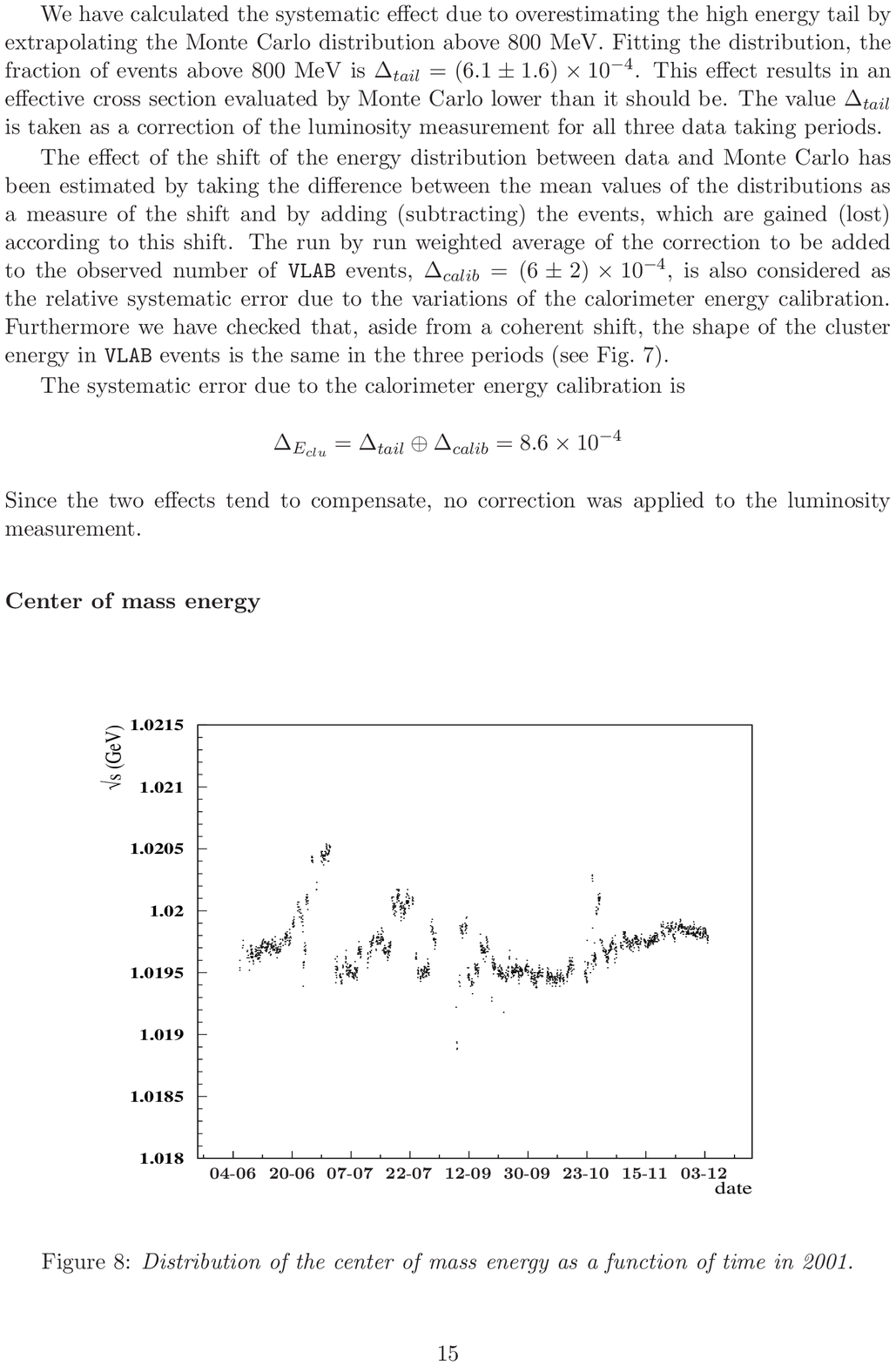}
}
\caption{Distribution of the
center of mass energy as a function of time in 2001.}
\label{fig:8} 
\end{center}
\end{figure}
The effective \texttt{VLAB} cross section is evaluated by Monte
Carlo at the average value of the center of mass energy,
$\sqrt{s}=1.0195\GeV$. To account for variations of the beam
energy, we corrected the luminosity measurement for the relative
change in cross section, $\Delta\mathcal{L}/\mathcal{L}_0=
-\Delta\sigma/\sigma_0$, where $\mathcal{L}_0$ is the luminosity
obtained with the nominal cross section $\sigma_0 = \sigma_{\rm
eff}(1.0195\GeV)$. The \texttt{VLAB} events energy scale was
calibrated with the well measured value of $M_\phi$~\cite{PhiMass}
during an energy scan~\cite{EnergyScan} around the
$\phi$ resonance. Figure~\ref{fig:8} shows as a
function of time the value of $\sqrt{s}$ measured from
\texttt{VLAB} events: variations in $\sqrt{s}$ never exceed a
few hundred keV, and single run corrections are always smaller
than 0.5\%. The average correction is 0.1\% and we consider this
value also as the systematic uncertainty for this effect.

\section{The \texttt{VLAB} cross section}
\label{sec:4}
\begin{figure}
\begin{center}
\resizebox{1\columnwidth}{!}{%
  \includegraphics{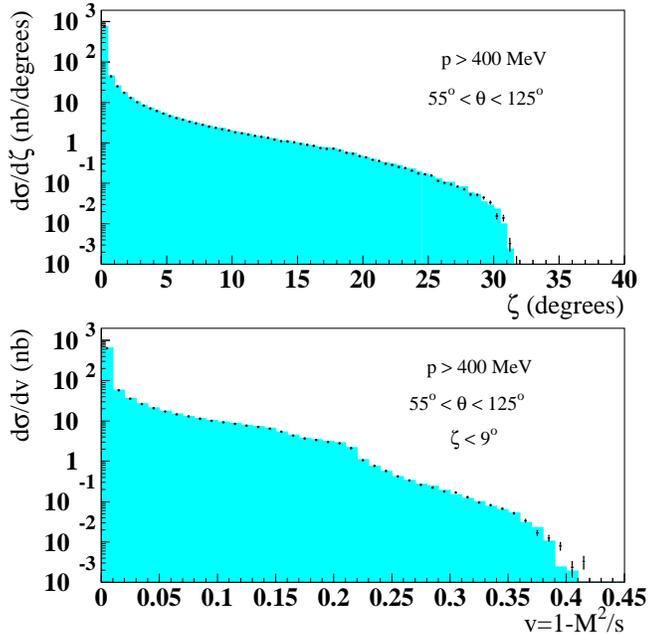}
}
\vspace*{-0.3cm}
\caption{Top: comparison between
\texttt{Babayaga} (histogram) and \texttt{Bhwide} (points)
for the differential cross section as function of the
acollinearity (after momentum and polar angle cuts). Bottom:
comparison between \texttt{Babayaga} and \texttt{Bhwide} for the
differential cross section as function of the missing energy
fraction, $v = 1 - M^2/s$ (after momentum, polar angle and
acollinearity cuts).}
\label{fig:9}
\end{center}
\end{figure}
The event generators \texttt{Babayaga}~\cite{CarloniCalame:2000pz,CarloniCalame:2001ny}
and \texttt{Bhagenf}~\cite{Drago:1997px}, developed
for the large angle Bhabha scattering at DA$\Phi$NE and based on the cross
section calculated in~\cite{Berends}, have been interfaced with
the detector simulation program \texttt{GEANFI}
\cite{Ambrosino:2004MC} for evaluating the effective cross section,
as well as for estimating the systematic uncertainties. After
applying the \texttt{VLAB} selection we find an agreement
better than $0.1\%$ between the cross sections calculated with the
two generators, including the event reconstruction efficiency:
\[ \mathtt{Babayaga} \quad \sigma_{\rm eff} = (431.0 \pm 0.3) \
 \mathrm{nb} \]
\[ \mathtt{Bhagenf} \quad \sigma_{\rm eff} = (430.7 \pm 0.3) \
 \mathrm{nb} \]
The error given in the above cross section
is due to the Monte Carlo statistics.
The systematic theoretical uncertainty claimed by the authors is
0.5\% in both cases. The radiative corrections due
to the treatment of initial and final state radiation
in \texttt{Bhagenf} and \texttt{Babayaga}
have been compared with two
other event generators: the \texttt{Bhwide} code~\cite{Jadach:1995nk}
developed for LEP and the \texttt{Mcgpj}
code~\cite{Fedotovich:2003ww} developed for VEPP-2M and
based on the cross section calculated in~\cite{Arbuzov:1997pj}.
Further details on the event generators and the application in the
analysis can be found in reference~\cite{KLOEnote}.
For this comparison, we applied the kinematic \texttt{VLAB} requirements on
the generated momenta and computed the \texttt{VLAB} cross sections for the four
generators, as shown in the table below, where errors are due to
Monte Carlo statistics. 
 \begin{center}
\vspace*{0.5cm}
 \begin{tabular}{lcc}
\hline\noalign{\smallskip}
MC code & ~ & $\sigma$ (nb) \\
\noalign{\smallskip}\hline\noalign{\smallskip}
\texttt{Bhagenf} & & $460.8 \pm 0.1$ \\
\texttt{Babayaga} & & $459.4 \pm 0.1$ \\
\texttt{Mcgpj} & & $457.4 \pm 0.1$ \\
\texttt{Bhwide} & & $456.2 \pm 0.1$ \\
\noalign{\smallskip}\hline
 \end{tabular}
\vspace*{0.5cm}
 \end{center}
These values are obtained without
considering detector smearing and loss effects and therefore the results
are considerably different from the effective \texttt{VLAB} cross
section presented before, where a full detector simulation was
performed. Moreover, contributions from the $\phi$ decay and vacuum
polarization effects are not applied, because they are the same
for all generators. 

The agreement among the four generators supports the systematic
uncertainty of 0.5\% quoted by the authors
of \texttt{Bhagenf} and \texttt{Babayaga}.

Moreover, we have compared the
differential distributions for the
acollinearity $\zeta$ and the missing energy fraction $v = 1 -
M_{ee}^2/s$, which are very sensitive to the difference in the treatment
of radiative effects. Also in this case we find good agreement as
can be seen in Figure~\ref{fig:9}.

\section{Results}
\label{sec:6}
The analysis refers to the data taken during the year 2001 for an
integrated luminosity of 141 pb$^{-1}$. All corrections and
systematic errors discussed above are summarized in
Table~\ref{tab:2}. Summing all errors in
quadrature, the relative experimental uncertainty for the
luminosity measurement using Bhabha scattering events
is $\delta\mathcal{L}_{\rm exp}/\mathcal{L}_{\rm exp}=0.3\%$.

\begin{table}
\caption{Summary of the corrections and
systematic errors in the measurement of the luminosity.}
\label{tab:2}
  \begin{center}
 \renewcommand{\arraystretch}{1}
 \setlength{\tabcolsep}{0.9mm}
   \begin{tabular}{|lcc|}
\hline\noalign{\smallskip}
~  & correction (\%) & systematic error (\%) \\
\noalign{\smallskip}\hline\noalign{\smallskip}
angular acceptance    & +0.25 & 0.25 \\
tracking              &  --   & 0.06 \\
clustering            & +0.14 & 0.11 \\
background            & -0.62 & 0.13 \\
cosmic veto           & +0.40 & --   \\
energy calibration    &  --   & 0.10 \\
center of mass energy & +0.10 & 0.10 \\
\noalign{\smallskip}\hline\noalign{\smallskip}
                      & +0.34 & 0.32 \\
\noalign{\smallskip}\hline
    \end{tabular}
 \end{center}
 \end{table}

Different event generators were used to evaluate the cross
section, the comparison shows good agreement in the distributions
of the variables used to select the events and in the value of
$\sigma_{\rm eff}$. The value of the effective \texttt{VLAB} cross
section has been calculated with the \texttt{Babayaga} event
generator that has been interfaced with the \texttt{GEANFI}
simulation program. 
We use a theoretical uncertainty of $0.5\%$, that is 
quoted by the authors of \texttt{Bhagenf} and
\texttt{Babayaga} (an improvement by more than a factor 2 is currently
in progress~\cite{CMCC06})
and it is confirmed by the comparison with other event
generators. 

The total error of the luminosity measurement is then
\[ \frac{\delta\mathcal{L}}{\mathcal{L}} =
   \frac{\delta\mathcal{L}_{\rm exp}}{\mathcal{L}_{\rm exp}} \oplus
   \frac{\delta\sigma_{\rm eff}}{\sigma_{\rm eff}} = 0.6 \% \]

\begin{acknowledgement}
\section{Acknowledgements}
We thank Carlo Michel Carloni
Calame, Stanislaw Jadach, Guido Montagna, Fulvio Piccinini
and Alexei Sibidanov
for the many fruitful discussions.
We thank the DA$\Phi$NE team for their efforts in maintaining low background running 
conditions and their collaboration during all data-taking. 
We want to thank our technical staff: 
G.F.Fortugno for his dedicated work to ensure an efficient operation of 
the KLOE Computing Center; 
M.Anelli for his continuous support to the gas system and the safety of
the
detector; 
A.Balla, M.Gatta, G.Corradi and G.Papalino for the maintenance of the
electronics;
M.Santoni, G.Paoluzzi and R.Rosellini for the general support to the
detector; 
C.Piscitelli for his help during major maintenance periods.
This work was supported in part by DOE grant DE-FG-02-97ER41027; 
by EURODAPHNE, contract FMRX-CT98-0169; 
by the German Federal Ministry of Education and Research (BMBF)
contract 06-KA-957; 
by Graduiertenkolleg
`H.E. Phys. and Part. Astrophys.' of Deutsche Forschungsgemeinschaft,
Contract No. GK 742; 
by INTAS, contracts 96-624, 99-37; 
and by the EU Integrated Infrastructure
Initiative HadronPhysics Project under contract number
RII3-CT-2004-506078.
\end{acknowledgement}

\end{document}